# Design considerations for a new generation of SiPMs with unprecedented timing resolution

**S. Enoch[a], A. Gola[b], P. Lecoq[c,d,1] and A. Rivetti[e]**

   a.   *Aix Marseille Univ, CNRS, Centrale Marseille, Institut Fresnel, Institut Marseille Imaging, Marseille, France*
   b. *Fondazione Bruno Kessler (FBK), Trento, Italy*
   c. *CERN, Geneva, Switzerland*
   d. *Polytechnic University of Valencia (UPV), Spain*
   e. *University of Torino, Italy*

   *E-mail* : paul.lecoq@cern.ch

ABSTRACT:

The potential of photon detectors to achieve precise timing information is of increasing importance in many domains, PET and CT scanners in medical imaging and particle physics detectors, amongst others. The goal to increase by an order of magnitude the sensitivity of PET scanners and to deliver, via time-of-flight (TOF), true space points for each event, as well as the constraints set by future particle accelerators require a further leap in time resolution of scintillator-based ionizing radiation detectors, reaching eventually a few picoseconds resolution for sub MeV energy deposits.

In spite of the impressive progress made in the last decade by several manufacturers, the Single Photon Time Resolution (SPTR) of SiPMs is still in the range of 70-120ps FWHM, whereas a value of 10ps or even less would be desirable. Such a step requires a break with traditional methods and the development of novel technologies.

The possibility of combining the extraordinary potential of nanophotonics with new approaches offered by modern microelectronics and 3D electronic integration opens novel perspectives for the development of a new generation of metamaterial-based SiPMs with unprecedented photodetection efficiency and timing resolution.

---

[1] Corresponding author.

**Contents**

# 1. Introduction:

The future generation of radiation detectors is ever more demanding on timing performance for a wide range of applications, such as particle identification in nuclear physics and high energy physics detectors, high resolution finely segmented hadronic calorimetry, precise event time tagging in high luminosity accelerators, time of flight (TOF) techniques for PET cameras and a number of photonic applications based on single photon detection.

Femtosecond laser sources can probe physical or biological processes with a high timing precision, and interesting developments in ultrafast scintillator heterostructures, allow converting X-rays and γ-rays into a bunch of prompt optical photons with a sub-nanosecond precision [1].

However, an important bottleneck remains at the photoconversion level, particularly when ultimate single photon detection precision is required. The recently launched 10ps Time-of-Flight PET challenge [2] has triggered several studies to identify the limits of state-of-the art technologies. In particular, a detailed study of the performance of many SiPMs from different producers has been conducted and recently reported in [3]. It was shown (Fig. 1) that, in spite of impressive progress in recent years, the SPTR (Single Photon Time Resolution), is in the range 70-120ps FWHM, which remains an important limitation for applications where 10ps timing resolution is desired.

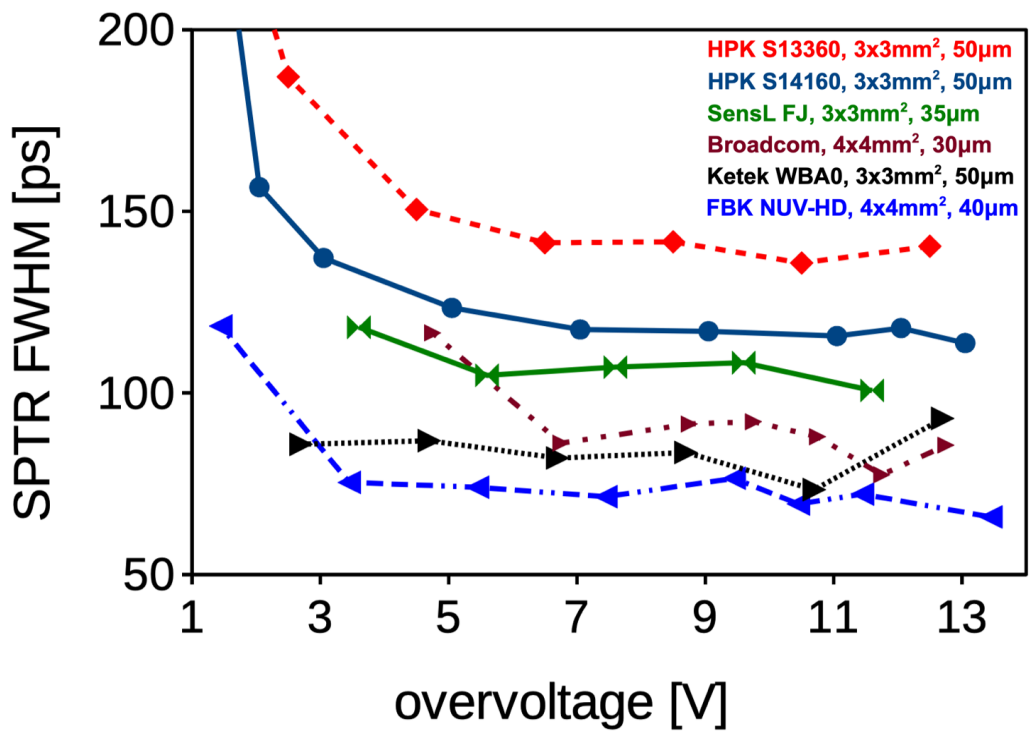

Fig. 1: SPTR of state-of-the-art SiPMs from different manufacturers as a function of the applied overvoltage. 42ps FWHM laser pulse width and electronic noise quadratically subtracted. From Ref [3]

A number of factors are limiting the time resolution of SiPMs, which can be classified in four main categories:

- The factors affecting the photostatistics, the time resolution being determined, in first approximation, by the inverse square root of the number of detected photons. SiPMs are an assembly of a large number of microcells (also called single photon avalanche diodes or SPADs) working in Geiger mode. These SPADs need to be electrically and optically isolated, which introduces dead space between them, where impinging photons cannot be detected. The photodetection efficiency (PDE) of a SiPM is the product of the fill factor (FF) of the SPADs on the surface of the SiPM times the quantum efficiency (QE) of each SPAD, and the avalanche triggering probability ($P_t$), i.e. the probability of generating an avalanche in the SPAD, once an electron-hole pair is generated inside the active area. FF, QE and $P_t$ are all smaller than 100%, which directly impacts the number of detected photons and therefore, the photostatistics. In spite of impressive progress in recent years, more than one third of the photons in the visible spectrum and as much as 80% in the UV and up to 90% in the near IR are on average not detected.

- The time to reach the detection threshold, and therefore the time jitter in the development of the avalanches in the SPADs. One source of this jitter is related to deviations from the electric field uniformity across a SPAD, as demonstrated in ref [4]. A second source is due to the variations of penetration depth of photons in the photodetector before it interacts with the silicon and produces and electron-hole pair, starting the avalanche. One part of these fluctuations is related to the stochastic process of the light-matter interaction. But another part results from the fact that the source of light is, in many cases (for instance if it comes from a scintillator), neither collimated nor monochromatic.

- The avalanche buildup statistics, caused by statistical fluctuations in the current growth in the SPAD after the first electron-hole pair is photo-generated [5, 6]

- The noise and capacitance of the SiPM, affecting the baseline and signal leading edge slope, with a direct impact on the time resolution

A large number of academic and industrial groups are working on different ways to reduce these different contributions to the time jitter. For instance, the introduction of microlenses in front of the SPADs allows to decrease the size, and therefore the capacitance of the SPADs, while reducing the influence of the FF and increasing the photostatistics [7]. However, while this approach brings significant improvements for a collimated source of light, it is inefficient for photons extracted from a scintillator due to their very large angular distribution. Major efforts are also devoted to reducing the crosstalk and after-pulses, to improving the design of the SPADs to minimize the influence of electric field inhomogeneities on the edges, to better control the jitter associated to the drift time of the carriers to the depleted region, and to reduce the SPAD capacitance and noise. In particular, charge focusing mechanisms in the drift region permit increasing the FF even for small cell sizes which are in general preferable for several reasons, including a lower correlated noise and an increased radiation hardness. Several of these developments are covered by patents, but interested readers can find useful information in recent papers and related references, for instance in [8, 9, 10, 11].

The progress in 3D microelectronics integration, as well as the fantastic potential of the emerging and rapidly expanding field of nano-photonics opens interesting perspectives for a performance breakthrough. There is an opportunity to capitalize on the huge efforts and impressive R&D investments deployed in the very hot domains of photovoltaic and imaging sensors for integrating nano-photonic features in microelectronic devices.

This paper describes an innovative concept of a Quantum Silicon Detector (QSD), a family of highly efficient and ultrafast single photon detector, integrating for the first time in a single compact device several features, as shown schematically in Fig. 2:

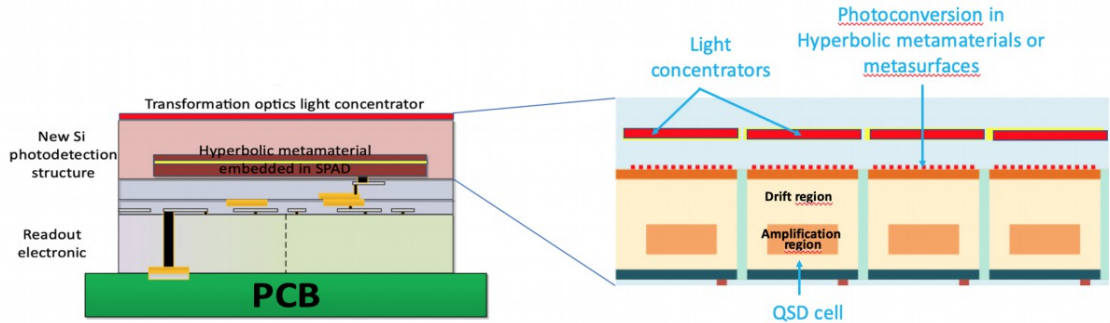

Fig. 2: Principle of the Quantum Silicon Detector

- Backside illuminated microcells, to enhance the Fill Factor and move the Back End of Line (BEOL) to the opposite side with respect to the light entrance window, thus providing an optimal surface to apply nanophotonics enhancements.

- Very small cell pitch of approximately 10μm, to reduce as much as possible detection non-idealities and, thus, preserve all the information carried by the light quanta. Indeed, such small cells will minimize both primary and correlated nose, even after irradiation, while extending significantly the QSD linear dynamic range (see Section 2).

- Transformation optics concentrators or hyperlenses with a high light concentration power over a large spectral range and angular acceptance, to guide photons to an innovative type of SPAD. This would provide an elegant way to improve the fill-factor and hence the photodetection efficiency compared to conventional SiPMs and to boost the PDE to values approaching 100%. For reference, current NUV-HD FBK SiPM technologies provide a FF close to 80% for the largest cell sizes of 40 $\mu$m, but this value is reduced to 55% for a 15$\mu$m cell [12]. Similar values are reported by other SiPM manufacturers with standard technologies.

- A very thin photo-detection layer to localize as well as possible the creation of electron-hole pairs and consequently reduce the time jitter in the development of the avalanches. Hyperbolic metamaterial layers can be designed to enhance at a well-defined depth of the QSD structure the electromagnetic energy density of the photons in the near field of these nanomaterials. This will locally increase the optical density of states and consequently the avalanche triggering probability and the quantum efficiency (QE), and reduce accordingly the time jitter. Considering that electrons are collected with a saturated drift velocity of about 0,1μm/ps in silicon and the penetration depth of the photons in silicon is a stochastic process following an exponential law, but is also wavelength and angular dependent, an ultimate target of 10ps SPTR requires converting the impinging photons in a charge drift layer not exceeding 1μm thickness.

- A suitable design of the dopant layers for a minimization of the time jitter and avalanche buildup jitter inside the SPAD [5, 6,13].

- A charge focusing mechanism, aiming at reaching high values of FF even in the case of small cells, i.e. 10$\mu$m or smaller. This feature works in conjunction with the concentrator

/ hyperlens, to provide the highest FF possible, very close to 100%, while simplifying the design of the concentrator/ hyperlens.
- Highly integrated front-end electronics readout, including time digitizers for each individual SPAD or for a very small group of SPADs in a 3D electronics architecture layout, reducing the noise and the capacitance of the photon detection chain. Indeed, as shown in ref. [14], in a typical SPTR measurement setup, the timing resolution is limited by the combination of detector output capacitance and the noise and bandwidth of the readout electronics. The larger the SiPM active area involved in the time measurement, the more important is this effect.

Combining even a subset of the characteristics mentioned above would provide a significant progress with respect to the state-of-the-art in specific application fields. Depending on the application domain, and in particular, on the wavelength range of interest, some of the proposed solutions will have more weight than others. For instance reading out the light from scintillators emitting in a broad energy range centered in the blue/near UV part of the spectrum with a very large angular distribution imposes different constraints to the photodetector than NIR Lidar applications with collimated light. In the former case the light concentrator will have to be specifically designed using non-resonant nanostructures to cope with a broad energy range centered in the blue/near UV part of the spectrum and a very large angular distribution of the incoming photons in contrast to Lidar applications with NIR monochromatic collimated light, for which more standard microlenses can be used. On the other hand, the importance of controlling the conversion depth of NIR Lidar photons will be more important than for scintillators applications in the near UV/blue part of the spectrum. The purpose of the QSD development is to combine the best technologies currently available in several fields, such as nanophotonics, SPAD technology and readout electronics, and to customize their detailed design and implementation in the photodetector to allow to get to an ideal photon detector, i.e. to a detector that preserves as much as possible all the information carried by the light quanta. Thus, such a detector must be able to detect close to 100% of the incoming photons, with excellent timing stamping and minimal noise, possibly even after irradiation, and negligible non-linearity and dead time. To achieve this goal, we believe that all the technologies listed above must be combined in a single device, i.e. in the QSD.

## 2. The Quantum Silicon Detector (QSD):

In the conceptualization of the QSD microcell, we start from the observation that, in recent years, two transformative technologies in integrated circuitry (IC) have revolutionized the field of silicon imaging sensors: backside illuminated devices (BSI) and 3D integration. These technologies have now matured enough to become industry standards in the field of consumer and professional imaging sensors, with Sony leading the field [15, 16, 17, 18]. Furthermore, 3D integration is a general trend in the silicon industry and, considering for example microprocessors, is considered by many observers as the only way to progress in line with Moore's Law [19, 20].

BSI allows increasing the fill factor of pixels compared to front-side illumination (FSI). At the same time, the interconnection layers, or the "*Back-End-Of-Line*" (BEOL), are moved from the light entrance window to the opposite side of the circuit, thus making way for an easier, more efficient, optical stack with enhanced QE and lower inter-pixel crosstalk [18]. Owing to these important changes, the pixel size of *imaging sensors* could be reduced to the current value of ~1μm [16]. BSI is also a pre-requisite to fully exploit the advantages of 3D integration techniques. 3D integration is motivated by the possibility of: (i) separately optimizing the sensor layer and the frontend or digital layers; (ii) adding functionality to the chip; (iii) decreasing the form-factor; (iv) increasing the fill-factor of the pixels; and (v) enabling flexible manufacturing processes.

First implementations on 3D-integrated CMOS imaging sensors (CIS) relied on through-silicon-vias (TSV), whereas more recent processes resort to hybrid bonding such as *Direct-Bond-Interconnects* (DBI).

Starting from these considerations, the QSD aims at transferring to the SiPM most of the transformative technologies that have revolutionized the CIS field. This requires evaluating how such technologies can be optimized specifically in the field of Geiger-mode detectors. The QSD cell is thus a backside illuminated microcell, organized in arrays with direct connection of each cell to the front-end electronics, thanks to advanced 3D integration techniques.

The QSD microcell will be operated above breakdown, in Geiger-mode, to provide single-photon sensitivity and has a size reduced by almost an order of magnitude compared to the SPAD size in conventional SiPM technologies, ranging between 5 $\mu$m and 10 $\mu$m. Gain is reduced, to a first approximation, proportionally to the area of the device, so that a reduction between one and two orders of magnitudes is expected, to a value of approximately $10^4$ or lower. Reduction of gain provides dramatic reduction of correlated noise, to first order proportional to the cell gain, which is a substantial contributor to the detector's overall performance, including timing resolution, photon number resolution and low-level triggering capability. Smaller cell capacitance also provides an ultra-fast recovery time constant $\tau_R$, even when using the traditional but very effective passive quenching scheme, because, to first approximation, $\tau_R$ is proportional to $C_T \cdot R_q$ ($C_T$ is the total cell capacitance and Rq is the value of the quenching resistor). A value of $\tau_R$ of a few nanoseconds is expected. A very fast recharge time, in combination with a cell density of up to 40 000 cells/mm$^2$, extends the linear dynamic range of the device, which allows the QSD to better approach an ideal photon detector and is of great importance for high-dynamic range energy spectroscopy applications [15]. A very fast recharge time also allows a significantly increased high-rate capability, as required by the next-generation of particle physics experiment at high luminosity colliders.

As a potential drawback, reducing the gain can be challenging for the design of the front-end electronics, to keep or even improve over the current level of timing performance. However, the QSD approach is to radically remove this limitation by connecting each cell directly to the front-end electronics through 3D interconnection techniques, as described in section 5. Based on simulations of the readout, even the smaller signal charge provided by the QSD cell is more than enough to guarantee the desired timing performance.

Attempts by two different groups to measure the SPTR of two single SPADs produced by FBK (10$\mu$m and 30$\mu$m diameter) with discrete electronics have shown that an intrinsic timing resolution of 21ps and 20ps FWHM respectively could be obtained for single photons, even at low overvoltage, in spite of the fact that these SPADs were not yet optimised for ultimate timing resolution [4, 37]. A target resolution of 10ps is therefore not out of reach with a careful optimisation of the main sources of time jitter, as proposed in this paper.

Most of the desirable features described in the previous paragraphs can be achieved by the significant reduction of the microcell size. However, as reported in [21, 22], there are technical limits to arbitrarily scaling down conventional SiPM microcells, due to the need of leaving extra space between the cells to ensure electrical isolation and prevent edge breakdown. This limitation can be reduced with the use of light concentrators / hyperlenses to focus the incident light to the very central part of the microcell, freeing the rest of its space for the necessary guard rings, as described in section 3. However, this approach might not allow reducing the cell pitch down to 5$\mu$m (or below) and the need for very high concentration power can complicate the design and implementation of the concentrators. For these reasons, and in addition to the light concentrators, the QSD cell design includes a charge focusing mechanism, as shown in Fig. 3(a), aiming at reaching a FF approaching 70%, even in the smallest cell sizes of 5$\mu$m. While the BSI SPAD cells with focusing of the charge have been proposed as early as 2006 [11], each proposed structure has different advantages and drawbacks for a given application and implementation. For example, the structure proposed in [11] requires electric field shaping rings, similar to the ones used in Silicon Drift Detectors. Because of them, it was not optimized for the highest cell density and is characterized by a minimum cell pitch of 130$\mu$m, which is not compatible with our goals.

To implement the proposed charge focusing mechanism the QSD cell is separated in two regions. The first one, located closer to the light entrance window, is the collection region, in which incoming light is adsorbed and converted to pairs of carriers. Because of the polarity of the junction, holes drift towards the light entrance window and do not initiate the avalanche. The electrons, which initiate the avalanches, drift and are focused, under the action of a suitably shaped electric field, obtained thanks to a proprietary device structure, to the second region, the multiplication region. Thickness of the collection region can be tuned, without much impact on charge collection properties, to ensure that > 90% of the visible or UV photons are absorbed in this region. The electric field in this region is optimized for fast charge collection, saturating, if possible the electron drift velocity while not reaching avalanche condition. The collection speed is also improved by the tiny lateral size of the cell. Trenches are used for electrical isolation and to reduce optical crosstalk between adjacent microcells. Thanks to the avoidance of the virtual guard ring in the collection region, the FF loss is determined by trenches size only, which can be reduced below 1$\mu$m, depending on the cell thickness.

As regards detection of NIR photons, in addition to increasing the collection region thickness, the BSI structure of the QSD is well-suited for the implementation of light trapping techniques aimed at increasing the optical path of the photons inside the device and the absorption probability. While a large number of possible implementations is described in literature (see for instance ref. [23]), the simplest implementation is a reflective surface on the front side of the QSD so that the NIR photons, impinging on the backside with normal incidence angle, double their optical path in the microcell. Assuming a thickness of the collection region of around 10$\mu$m in this case, and considering that the absorption coefficient of silicon for 905nm IR photons (used in LIDARs) is 35$\mu$m, doubling the optical path corresponds to approximately doubling the detection efficiency. Other solutions, based on hyperbolic metamaterial layers are presented in section 4 of this paper.

We also note that Back End of Line (BEOL) structures are mainly placed on the front side of the QSD cell and, thus, the light entrance window, located on the backside, is free from any other structures, save for the presence of trenches. This condition is ideal to fabricate complex optical stacks, such as those required by the light concentrators.

Considering now the multiplication region, thanks to the efficient charge focusing mechanism, the high electric field, necessary to initiate avalanche breakdown, is present only in a volume, $V_{av}$, with base $A_{av}$ and height $Z_{av}$ as shown in Fig. 3. $A_{av}$ is significantly smaller than the cell area, leaving sufficient space at the cell edge for guard ring structures. Depending on the QSD cell size and layout variant, the avalanche area has a diameter ranging between 2.5$\mu$m and 7.5$\mu$m. We note that, in several cases, a majority fraction of the SiPM dark counts can be attributed to carriers generated in the high-field region, as the carrier generation rate follows the field-enhanced Shockley–Read–Hall (SRH) statistics. For example, NUV-HD SiPMs fabricated at FBK show an activation energy of their dark count rate (DCR) of 0.42eV near room temperature, while their low-field counterpart, the NUV-HD-Cryo, featuring lower electric field in the high-field region, increase this value to 0.48eV [11]. These measurements underline the importance of the high-field volume in determining the value of the DCR in a given device. The fact that the focusing mechanism allows having a much smaller high-field area (and volume) in the microcell, compared to standard SiPM technologies, without reducing the FF, is a clear advantage.

We expect interesting research, potentially yielding to relevant improvements on the avalanche build-up and quenching characteristics in the QSD, considering that the gain is much reduced and the avalanche is confined into a much smaller area, $A_{AV}$, comparted to the cells of a more traditional SiPM, in which the high-field region width is usually tens of microns. For reference, studies on the avalanche build-up characteristics also related to the avalanche lateral spread are reported in [5, 13], for single SPADs without passive quenching. In [24], the avalanche diameter in a SiPM with passive quenching is estimated at $10\mu m$.

SiPMs with small cells, characterized by small gain and fast recovery time, are also less sensitive to the effects of radiation damage and are thus preferable when the requirements on radiation hardness are especially demanding, such as, for example, in particle physics experiments at high luminosity colliders. The QSD is an ideal detector for such applications, thanks to the ultra-low gain and ultra-fast recovery time. Moreover, we expect that the most relevant effect of irradiation-induced bulk-damage on the QSD noise, i.e. increase of DCR, is caused by defects created in the high-field, multiplication region, through trap-assisted tunneling or the Poole–Frenkel effect. As mentioned above, the QSD cell structure is particularly advantageous, because the high-field volume, which is the most sensitive to radiation damage, is significantly smaller than the charge detection and collection volume thanks to the charge focusing mechanism.

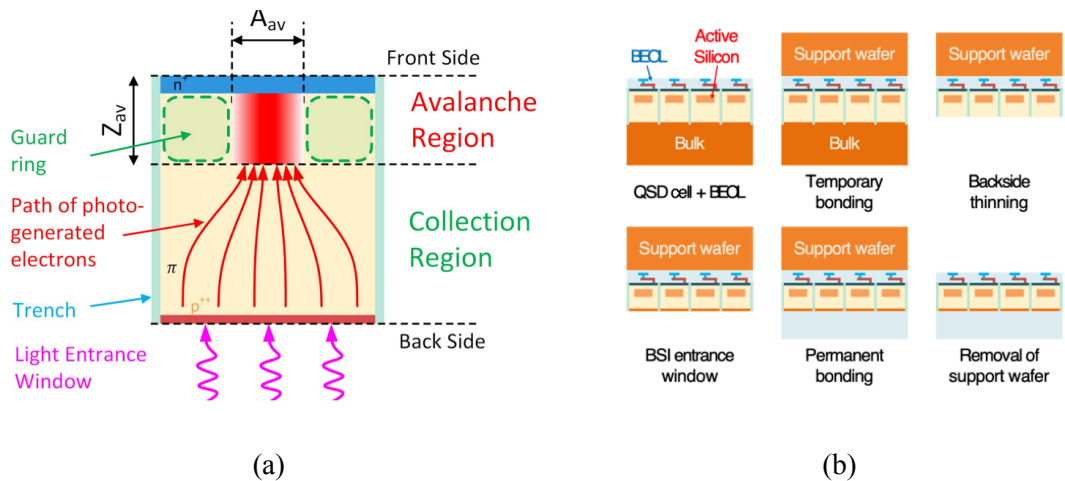

Fig.3: (a) Schematic representation of the working principle of the QSD cell; (b) BSI process flow

QSD sensors will be fabricated on epitaxial silicon. Multiplication regions and isolating structures will be fabricated on the front side, along with back-end-of-line (BEOL). Then, we will apply a relatively standard BSI processing, schematically represented in Fig.3(b). Specific research effort will be dedicated to the formation of the light entrance window on the backside, which needs to provide good internal QE also at short wavelengths for which photon absorption is very superficial (for reference, the absorption depth of photons at 360nm and below is approximately 10nm). Baseline solution will be low-energy ion implantation and followed by laser annealing, properly tuned to ensure good QE at shorter wavelengths. Molecular Beam Epitaxy (MBE) will also be investigated to produce an effective entrance window with reduced surface recombination velocity and good charge collection efficiency. To extend sensitivity to VUV photons (178nm and below), alternative 3D integration process flows, compared to the one reported in Fig.3(b) will also be investigated. In this case, we will remove the borosilicate glass, which is normally employed as a permanent support wafer and is transparent to visible and near-visible photons but not to VUV ones. As an alternative solution, we will rely on the front-end electronics wafer for mechanical support (active interposer concept).

Summing up, from our analysis, we conclude that techniques such as BSI and 3D integration provide very relevant advantages for the QSD and are, thus, the most promising options to be pursued in the design of this next-generation detector. The advantages include: (i) better approximation of an ideal photon detector, trough the improvement of almost all electrooptical parameters, such as reduction of noise, increased linear dynamic range, rate capability and radiation hardness; (ii) easier and more effective integration with the nanophotonics-based optical stacks described in the following sections; (iii) more efficient interconnection to the readout electronics, which allows optimal extraction of the timing information from the QSD cell and decouples the size of the QSD active area from its timing resolution. However, to fully exploit the advantages provided by the BSI, it is necessary to completely redesign the structure of the cell, based on a concept of double optical and electrical focusing mechanism.

## 3. Increasing the fill factor with nano-structured light concentrators

As mentioned above, in modern SiPMs, the active area of SPADs occupies between 50 to 80 percent of the surface, depending on the SPAD size and density.

Light concentrators on the surface of the QSD to guide photons onto the sensitive part of QSD microcells, regardless of the size of structures isolating adjacent microcells, would yield an effective fill-factor of nearly 100% of the QSD with a large impact on the photostatistics.

In order to reduce the effect of the filling factor, microlens arrays have been proposed. Most often they are made of polymers with a moderate optical index [25, 26, 27]. As a consequence, the focal lengths remain quite large, which poses a number of problems such as effects of non-collimated light, ie. sensitivity to incidence. In addition, these microlens arrays require of course to create non-planar surfaces. Recently, arrays of diffractive microlenses have been implemented on SPADs and have shown their ability to improve the sensitivity in the near-infrared [28]. The authors report a measured sensitivity improvement up to a factor 5.3 at 850nm for SPADs having a pitch of 21.6μm and a bare fill-factor of 5%. If this is a viable approach for LIDAR applications, it is unfortunately not applicable for non-collimated light sources, as for scintillator readout applications.

Other possibilities in this case can be provided by nano-photonics, repeatedly employed in the development of lens solutions over the last two decades. Nano-photonics has become synonymous with many breakthroughs, e.g. in plasmonics, nano-antennas and metasurfaces, and to control optical wave guiding unattainable with traditional optical techniques. One possible approach is based on transformation optics that has been highly publicized for its application for invisibility cloaks. Transformation optics has been proposed in an article in the journal Science entitled "controlling electromagnetic field" [29].

In this context, gradient index (GRIN) lenses can be seen as transformational optics objects [30]. Indeed, the first transformation-based design, which exposed the underlying physics of the Maxwell fish eye (imagined over 150 years ago) is due to the mathematician Luneburg, who derived the refractive index of this GRIN lens from the stereographic projection of a sphere on a plane. Recently, this transformation optics approach has been revisited by various authors [31, 32]. Authors of this paper and their collaborators have recently demonstrated a CMOS-compatible all-dielectric metalens with a 1.15 numerical aperture [33], which can be considered, for the time being, as an affective baseline solution for scintillator readout applications.

Another approach is based on the so-called hyperlens structures [34, 35] based on hyperbolic metamaterials (HMM). They are able to image beyond the diffraction limit and magnifying hyperlenses have been manufactured using tapered wire-arrays [36] or conformed thin films [35]. Similar structures can be used in the opposite way: to concentrate light instead of magnifying images, but with less demanding requirements with respect to imaging, as we are considering here photodetectors as light collectors only and not as imaging devices. The great advantage of hyperlenses over more standard microlenses including GRIN is their ultrashort focal length, allowing an important concentration factor even at large angular distribution and spectral range of the incoming photons. However, their integration in CMOS technology remains to be demonstrated.

Moreover, in addition to compensate for the reduced FF, light concentrators will cause carrier generation to happen always in the central region of QSD microcell active area, where the electric field is more homogeneous, thus improving timing performance, in terms of SPTR, as reported in [4, 37]. By concentrating the photogeneration of incident light in the central area of the microcell and at an optimal depth, a QSD structure integrating such nano-structured light concentrators could collect and detect nearly 100% of the photons.

## 4. To go further increasing the quantum efficiency with nanophotonics.

One source of time jitter in SiPMs coupled to scintillating crystals is related to the variation of the penetration depth of the photons before converting to an electron-hole pair, as a function of its wavelength (particularly important for Lidar NIR photon detection) and incident angle (for scintillator applications, although with a smaller impact because of the shorter scintillation wavelength and absorption depth of the majority of scintillators). Solutions should preserve a sufficient bandwidth, particularly for scintillators. This requirement excludes all the solutions based on excitation of highly resonant modes, such as photonic crystals cavities for example.

We are considering several approaches inspired from recent developments on hyperbolic metamaterials, near zero index metamaterials or metasurfaces to reduce the photodetection layer to a few hundred nm only (at saturated drift velocity in silicon, electron travel 100nm in about 1ps) and to strengthen the photon-electron coupling field in this thin layer in order to reach a quantum efficiency approaching 100%. The choice will be mostly determined by engineering considerations related to potential difficulties and cost effectiveness for implementing these structures in the QSD with standard microelectronic processes.

One of the most promising property of hyperbolic metamaterials (HMMs) is their promise for a large photonic density of states (PDOS) over a relatively large bandwidth [38, 39]. Indeed, most of the nanophotonic structures based on electromagnetic resonances that exhibit large density of states are intrinsically narrow band. Among these structures the most studied are those supporting surface plasmon polaritons [40], nanoparticles, nanoantennas, photonic crystal cavities [41] and micro-cavities. [42].

Two other types of metamaterials can also be considered: near zero index metamaterials (NZIMs) and metasurfaces. Even if these structures are considered a priori as narrow band, efforts from several researchers have succeeded in widening the bandpass. One of the very first article on NZIMs demonstrated that they can be used to control the interaction field between silicon atoms and photons passing nearby [43].

In the field of photovoltaics, several examples have shown that nanophotonics can achieve high conversion efficiencies in very thin silicon layers, a problem similar to the one to be solved to suppress the important source of time jitter in SiPMs related to the variable photon conversion depth [44, 45]. Although it is not the purpose of this paper to describe in detail the studies carried out in this field, we can cite the work carried out by Chen et al. [23] in which they have reported a 205nm-thick GaAs solar cell with a certified efficiency of 19.9% based on broadband light trapping achieved with multiple overlapping resonances. A similar strategy based on grating coupled-hyperbolic metamaterials has been used to develop plasmonic sensor with a record figure of merit for this class of sensors [23]. These very promising results and the extensive developments of these techniques in the domain of photovoltaics are pushing us to implement a vertically integrated hot-electron conversion device based on metamaterials that can efficiently couple plasmonic excitations into electron flows. It has been shown that such an approach is leading to a theoretical quantum efficiency approaching the physical limit [46]. The possibility to limit the conversion region to a thickness of a few hundreds of nanometers only would significantly reduce the jitter introduced by carrier drift time and minimize the influence of the photon incidence angle and wavelength, particularly in the infrared region.

Integrated in the QSD structure, HMMs have the potential to increase to nearly 100% the probability for a photon to trigger an avalanche and to suppress an important source of SiPM time jitter related to the variation of the carrier generation depth in silicon.

## 5. Highly integrated, low noise, low power consumption and ultrafast electronics

The QSD array will be readout by an application specific integrated circuit (ASIC). The small cell size and the target time resolution impose also for this part of the system an improvement with respect to the state of the art. Today, advanced photon TOF systems based on multi-tier approach, feature, as an example, a pixel size of 19.8μm x 19.8μm, a time resolution of 60ps. and a fill factor of 31% (see for instance [47] and references therein). The functionality of the readout electronics can be partitioned into four main tasks: signal amplification and discrimination, time-to-digital conversion (TDC), online data processing and data transmission. The signal amplification and discrimination must be performed at the level of the individual QSD cell. Traditional SPADs offer a large signal. In case each SPAD is read individually, the input stage can take the form of a simple inverter. In the QSD approach, the signal amplitude is treaded with speed and uniformity, so that the sensor offers a response that minimises signal shape variations from event to event. If the signal shape can be kept very stable, the reduction in gain is not expected to be a show-stopper. A fast signal with a total charge of $10^4$ electrons can still be easily detected by a fairly simple input stage with low jitter. Fig. 4 shows the input stage used in the preliminary study, based on a current comparator followed by an inverter. A total input capacitance of 50fF has been assumed. Fig. 5 shows the circuit response for a signal conveying a total charge of 2fC in 1 ns and a comparator threshold of 100nA. A clear digital pulse is produced at the output. The intermediate node at the inverter input is shown in the inset. In this point, the simulated r.m.s noise is 1.3mV, while the signal slope id $2.5*10^8$ V/s, which results in a random jitter of 5.2ps. From a random jitter standpoint, very good performance can thus be achieved even with small signals, while the ultimate timing performance will be dictated by signal shape fluctuations.

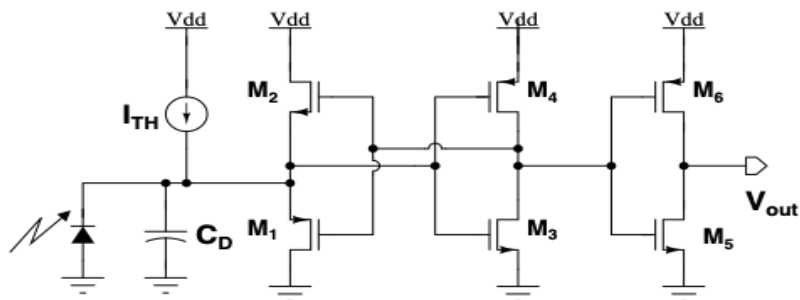

Fig.4 Schematic of the input stage used in simulations

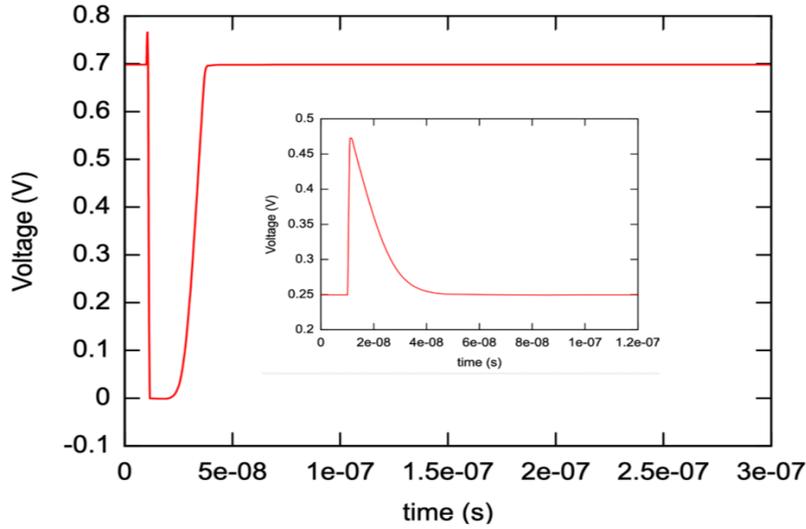

Fig.5 Input stage response. The input of the last inverter is shown in the inset.

The direct connection between the cell and the input stage of the front-end electronics is mandatory to minimise the parasitic capacitance and hence the timing jitter. Additionally, in applications like high resolution imaging it is necessary to store the address of the firing cell to preserve the spatial resolution. Therefore, the layout of the chip must closely match the QSD microcell arrangement. The connection can be provided by ultra-fine pitch bonding and Cu-Cu connection techniques, like the ones shown in Fig. 6 [48, 49]. In order to cope with the very small pixel size the use of a small feature size, such a 14nm or even smaller, is required. As it is often done in a 3D-SiPM, the TDC can be shared between several pixels in a given region. To match the 10ps resolution goal, the Least Significant Bit (LSB) of the TDC should not be larger than 5ps, which results in a quantisation noise of 1.5ps rms. In this way, the TDC gives a negligible contribution to the overall time resolution. The adoption of TDCs in an All-Digital PLL (ADPLL) to replace the analogue phase detectors has stimulated vigorous studies in the domain of ultra-high resolution TDCs and several converters with ps grade resolution have been reported [50, 51]. The research is now gearing towards TDCs with sub-ps resolution. However, in case of the QSD detectors, the challenge comes from the need to implement such very high-resolution converters with unprecedented density. The multi-tier 3D integration of SiPMs with an unprecedented degree of signal processing is still in its infancy and much progress is to be expected in the coming years [47]. This is why we would like to follow this trend in the QSD design.

High-speed serializers allow today to transmit up to 64 Gbit/s over a single differential line [52]. Nevertheless, even using the most aggressively scaled technologies, the power consumption of such a transmitter is of the order of 200mW and ultimately limits the amount of data that can be sent off-chip. In several domains, such as Positron Emission Tomography or Lidars to be used in autonomous cars, each event can produce a large number of photons, but it is not necessary to preserve the space resolution provided by the single QSD cell. An on-chip, real-time data reduction becomes thus feasible. This can be obtained by grouping the pixels into regions or macro-pixels. In each macro-pixel, a digital processing unit selects the time-stamps of only the first few photons which define the time-resolution, while the overall number of photons is counted to measure the total energy deposited by the event. Such an approach can reduce the amount of transmitted data by at least one or two orders of magnitude, depending on the total number of photons that are associated to a given event. In the optimal case, the number of microcells forming a macro-pixel is configurable, so that the same readout can serve different applications, each requiring its optimal space resolution. The need of a flexible, high performance and low power logic together with ultra-fast serialisers are other compelling factors that call for the use of a technology node with a feature size as small as possible. A process like the 14nm FinFET thus offers an interesting perspective [53]. An alternative approach for the readout electronics can be the use of a multi-tier ASIC implemented with 3D integration technologies [49]. By packing more independent chips in a vertical stack, a multi-tier approach allows to put more functionality in the same area, but the design complexity and costs make a simpler 2D implementation preferable.

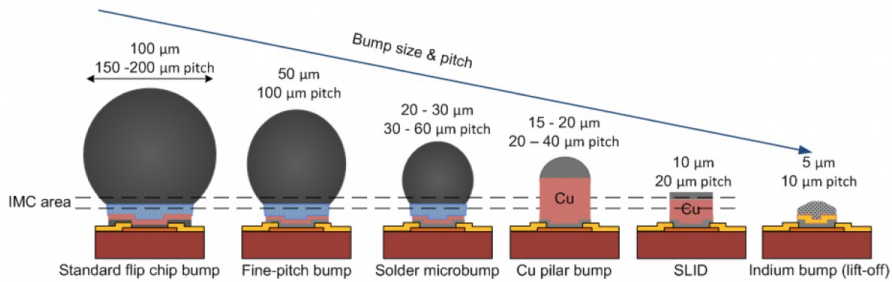

Fig. 6: Evolution of bump size and pitch. Courtesy of T. Tick and S. Vahanen

## 6. Conclusions

In order to meet the increasing demand for detecting photons with a timing resolution in the picosecond range for a number of applications, a radically new approach needs to be adopted in the design of a new generation of photodetectors. Efforts for developing superconductive nanowires are certainly going in this direction. [54]. However, their implantation in highly integrated multichannel devices and their cryogenic working temperature remain issues for complex multichannel detectors, such as PET scanners for instance. The alternative QSD approach proposed in this paper, based on room temperature solid state photodetectors, such as SiPMs, seems more realistic on a medium term and can be implemented is several steps, each one improving the timing performance. Recent years have seen impressive advances in the domains of micro-electronics, nanotechnologies and nanophotonics, which open the way to paradigm shifts in the design of this new class of solid-state photodetectors, intimately combining the three technologies.

Finally, we point out that not all the features of the QSD cell structure are disclosed in this work, because of patent applications pending. On the other hand, the purpose of the paper is to document the roadmap that we envision for the QSDs, which is based on several components, ranging from nanophotonics structures to advanced front-end, and also including a heavily modified microcell whose characteristics have been analyzed, in general terms, in the description above.

**Acknowledgement**
This concept was initiated in the frame of the ERC Advanced Grant TICAL (grant agreement No 338953) funded by the European Research Council. It is now partly being developed in the frame of the EU funded ATTRACT PHOTOQUANT project (Grant # 777222).